\documentclass[reprint,
superscriptaddress,	    
showpacs,                
amsmath, amssymb,        
aps, pra]{revtex4-1}	
\usepackage[greek, russian, spanish, english]{babel}
\usepackage{graphicx}    
\usepackage{dcolumn}     
\usepackage{bm}	         
\usepackage{xcolor}
\usepackage{amsmath} 
\usepackage{hyperref}    
\hypersetup{colorlinks=true, citecolor=blue, urlcolor=blue, linkcolor=blue}
\usepackage{subfigure}
\usepackage{ragged2e}
\usepackage{caption}
\begin{document}

\title{General phase diagram features of superradiant phase transitions}

\author{Wen Zhao}	    
\affiliation{Hunan Key Laboratory for Micro-Nano Energy Materials and Devices\\ and School of
Physics and Optoelectronics, Xiangtan University, Hunan 411105, China}
\author{Junlong Tian}
\affiliation{Department of Electronic Science, College of Big Data and Information Engineering, Guizhou University, Guiyang 550025, China}
\author{Jie Peng}	    
\email{jpeng@xtu.edu.cn}
\affiliation{Hunan Key Laboratory for Micro-Nano Energy Materials and Devices\\ and School of
Physics and Optoelectronics, Xiangtan University, Hunan 411105, China}
\date{\today}	        

\begin{abstract}
Various light-matter interactions lead to diverse phase diagram structures in superradiant phase transition (SPT) studies. Such systems consist of multiqubit and multimode with anisotropic couplings, one- and two-photon interactions, Stark shifts, inter-cavity hoppings, qubit-qubit interactions and so on. We find a general phase diagram  feature that the origin is in normal phase (NP) and SPT happens only once along the radial direction of a chosen coupling parameter vector with the mean-field method at finite temperature. We can calculate the phase boundary and SPT properties by a concise method. We illustrate it with specific models and find SPT can be achieved in strong coupling regime by means of multimode collective behavior. We also find the disorder will shift the phase boundary. These general features  facilitate SPT studies and their diverse applications.
\end{abstract}
\maketitle

\section{\label{sec:level1}Introduction}

The collective behaviors of a large number of atoms interacting with a cavity field unveiled by Dicke \citep{Dicke1954} leads to the superradiance. In recent decades, the investigation of the intriguing phenomenon in light-matter systems has attracted a great deal of attention. The so-called Dicke model, which describes a collection of qubits interacting with a single-mode cavity field, is predicted to exhibit a SPT between a NP and a superradiant phase (SP) at finite \citep{Hepp1973_1, Wang1973} or zero \citep{Emary2003} temperature in the thermodynamic limit. The SP in a quantum SPT features an acquisition of a macroscopic population of coherent photons for the cavity field in the ground state. Recently, a remarkable advance is the idea that the SPT can also occur in few-body light-matter systems \citep{Bakemeier2012, Ashhab2013, Hwang2015, Hwang2016}. The quantum Rabi model (QRM), which corresponds to the single-qubit version of the Dicke model, exhibits an equivalent SPT in an another limit, where the ratio of the qubit frequency to the field frequency approaches infinity, as demonstrated by Hwang \emph{et al.} \citep{Hwang2015}. It can be considered as the alternative of the thermodynamic limit. Such SPT holds significant promise for critical quantum metrology \citep{simon2020,Garbe_2022,ricard2022}. In addition, a general framework of universality of the SPTs for the generalizations of the QRM with the number of qubits varying from finity to infinity was established by Liu \emph{et al.} \citep{Liu2017}.

The recent experimental achievements of ultrastrong-coupling regime \citep{Forn2010, Niemczyk2010, Ballester2012} or even the deep strong coupling regime \citep{Bayer2017, Casanova2010, Youshihara2017} provides a foundation for the experimental observation of SPT \citep{Zhangxt2021,Cai2021}. The theoretical scheme \citep{ Huang2022} and the experimental control \citep{Baumann2010, DeBernardis2018} in the cavity quantum electrodynamic (cavity QED) systems have been intensively discussed. In addition, experimental platforms based on cold atoms \citep{Michelle2016}, superconducting circuits \citep{ Zhangyw2014}, quantum dots \citep{Petru2016}, trapped ions \citep{Devoe1996, Puebla2017, Gambetta2019, Cai2021}, Fermi gas \citep{Chen2015, Zhangxt2021}, and terahertz metamaterials \citep{Bayer2017} are platforms for realizing SPTs. 

Tremendous interest is focused on the SPTs of the generalizations of the QRM or Dicke model, such as anisotropy \citep{Liu2017, Baksic2014, Xie2020, yanzhi2020,Jiangxd2021}, Rabi triangle with artificial magnetic fields \citep{ZhangYY2021,ZhangYY2022}, nonlinear coupling process \citep{ Duan2016, Garbe2017, Chen2018, Cong2019,Chen2024}, inter-cavity hopping \citep{Greentree2006, Lei2008, Schiro2012, Schiro2013, Wangym2020}, multimode extension \citep{Alderete2016, Shen2021, Soldati2021}, qubit-qubit interaction \citep{Peng2019,Devi2020}, and so on \citep{Larson_2017, Xie2019,  Cui2019, Ying2020, Shu2022, hwang2022,xiong2023,LeBoit2023,lv2024,yangluo2025}. The competition between these various coupling processes leads to rich phase diagrams. A general feature was found in Ref.~\citep{Peng2019} in single-mode and dipolar-coupling case. Additionally, Felicetti \emph{et al.} \citep{Felicetti2020} have proven that the emergence of the SP is the consequence of the ultrastrong coupling limit in a wide class of models with bounded and unbounded operators. Thus, it is natural to raise a question on how a light-matter system enters the SP region from the NP region under the interplay of various coupling processes in the coupling parameter space. 

In this paper, we find a universal phase diagram feature for a general light-matter model with multiqubit, multimode and a variety of coupling processes, such as anisotropic couplings, linear and nonlinear qubit-photon couplings, inter-cavity hopping, and nonlinear Stark couplings. Since the mean-field method has been prove to be valid \citep{Peng2019ar}, we find the rescaled mean photon number is determined by the global minimum of a Landau potential (GMLP), located at the origin for NP and a nonzero point for SP. By analyzing the structure of the Landau potential, we find the system is naturally in NP at the origin of a chosen coupling parameter space, and enters SP region as parameter vector $\vec{\lambda}$ grows radially, and always stays there afterwards. The phase boundary and the rescaled mean photon number there can be obtained by a concise method, so that the SPT properties (first-order, second-order, or none) can be determined. We illustrate this with the multimode Dicke model with both one- and two-photon terms, the Rabi-Stark-Hubbard model, and the anisotropic Rabi-Stark model. Interestingly, the SPT condition for coupling $\lambda$ in the single mode case becomes correspondingly $\vert \vec{\lambda}\vert=\vert(\lambda_1,\lambda_2,\ldots)\vert$ for the multimode case, so the coupling strength for each mode can be much reduced, even to the strong coupling regime, where superradiance is achieved in a cooperative way. We also find the disorder will shift the phase boundary.
 
The paper is organized as follows. In Sec.~\ref{sec:level2}, we give a study on SPT of general light-matter systems and their general phase diagram features. We also give a concise way to calculate the phase boundary and SPT properties. We illustrate it with the multimode Dicke model with both one- and two-photon interactions in Sec.~\ref{sec:level3}, the Rabi-Stark-Hubbard model in Sec.~\ref{sec:level4}, and the anisotropic Rabi-Stark model in Sec.~\ref{sec:level5}, respectively. Sec.~\ref{sec:level6} gives the conclusions.

\section{\label{sec:level2}General features of phase diagram}

We aim to unveil the phase diagram feature of general light-matter systems: 
\begin{eqnarray}
H&=&\sum^{M}_{i}\omega_i{a^{\dag}_ia_i}+\sum^{N}_{j}{A^{\dag}_jA_j}+\sum^{M}_{i}{\sum^{N}_{j}{a^{\dag}_{i}a_{i}A^{\dag}_{j}A_{j}}}\nonumber\\
& &+
\sum^{M}_{i}{\sum^{N}_{j}{[\alpha_{ij}(a_i A^{\dag}_j\!+\! a^{\dag}_iA_j)\!+\! \beta_{ij}(a_i A_j\!+\! a^{\dag}_iA^{\dag}_j)]}}\nonumber\\
& &+
\sum^{M}_{i}{\sum^{N}_{j}{(a^2_i+a^{\dag 2}_i)(A_j+A^{\dag}_j)}}\nonumber\\
& &+
\sum^{M}_{i}{J_i(a_i+a^{\dag}_{i})(a_{i+1}+a^{\dag}_{i+1})}\nonumber\\
& &+
\sum^{N}_{j,k(j\neq k)}{(A_j+A^{\dag}_{j})(A_{k}+A^{\dag}_{k})}.
\label{eq:one}
\end{eqnarray} 
where $A_{j,k}$ ($A^{\dag}_{j,k}$) are atomic annihilation (creation) operators with coupling strengths included. $a_i$ ($a^{\dag}_i$) represents the  photonic annihilation (creation) operators. Here, $M$ denotes the number of the field modes or the cascaded cavities while $N$ is the number of the qubits. $\omega_i$ is the frequency of the $i$-th mode. $J_i$ is the hopping between $i$ and $(i+1)$-th mode. The dimensionless constants $\alpha_{ij}$ and $\beta_{ij}$ represent the anisotropic coupling coefficients. Other coupling constant is included in $A_{j,k}$ and $A^{\dag}_{j,k}$. Clearly, the Hamiltonian in Eq.~\eqref{eq:one} covers a wide class of specific models such as the multimode Dicke model with both one- and two-photon terms, the Rabi-Stark-Hubbard model, and the anisotropic Rabi-Stark model. 

To investigate its thermodynamic properties at equilibrium, we study the  canonical partition function $Z=\rm{Tr}e^{-\beta H}$, where $\beta=1/k_B T$. It can be expanded in the Glauber's coherent state basis as
\begin{eqnarray}\label{zap}
Z = \int \frac{\prod_{i=1}^Md^2\alpha_i}{\pi^M}
\, \rm{Tr_q}[   \langle\prod_{i=1}^M\alpha_i| e^{-\beta H} | \prod_{i=1}^M\alpha_i \rangle ].   
\end{eqnarray}
It has been proven the mean field approximation is valid in thermodynamic or classical oscillator limit \cite{Peng2019,Peng2019ar}, so we apply the mean field method proposed in Ref.~\citep{Wang1973} to obtain $\langle\prod_{i=1}^M\alpha_i| e^{-\beta H} | \prod_{i=1}^M\alpha_i \rangle=\langle\prod_{i=1}^M\alpha_i| e^{-\beta H(a_i\rightarrow\alpha_i,a^\dag_i\rightarrow \alpha^*_i)} | \prod_{i=1}^M\alpha_i \rangle $.
Therefore
\begin{eqnarray}\label{z0}
Z =  \int \frac{\prod_{i=1}^Md^2\alpha_i}{\pi^M}e^{-\beta \sum \omega_i|\alpha_i|^2}
\, \rm{Tr_q}[ \langle\prod_{i=1}^M\alpha_i| e^{-\beta H'} | \prod_{i=1}^M\alpha_i \rangle],  
\end{eqnarray}
where $H'=H-\sum_i\omega_ia_i^\dagger a_i$. Obviously, $\rm{Tr_q}[  \langle\prod_{i=1}^M\alpha_i| e^{-\beta H'} | \prod_{i=1}^M\alpha_i \rangle ]$ is a function of $\lambda_ix_i$ and $\lambda_{i'}y_i$, where $\alpha_i=x_i+iy_i$. $\rm{d^2}\alpha_i=\rm{d}x_i\rm{d}y_i$. $\lambda_i$ and $\lambda_{i'}$ are some coupling constant. 

To investigate the SPT, we rescale the mean photon number by $\eta_i=N\Omega/\omega_i$, where $\Omega$ is a positive generalized mean of qubit energies $\{\Omega_j \}$, so that $<a_i^\dagger a_i>/\eta_i$ is zero for NP and finite for SP under thermodynamic ($N\to\infty$) or classical oscillator limit ($\Omega/\omega_i\to\infty$). We redefine $u_i=x_i/\sqrt{\eta_i}$ and $v_i=y_i/\sqrt{\eta_i}$, and reduce Eq. \eqref{z0} to
\begin{eqnarray}\label{z1}
Z = \left( \prod_{i=1}^{M} \eta_i \right) \pi^{-M} \int e^{-\beta N \Omega \phi(\vec{z})} d\vec{z}, 
\end{eqnarray}
where 
\begin{equation}\label{landaup}
\phi(\vec{z})=|\vec{z}|^2+\mu(\vec{\lambda}\otimes\vec{z}).
\end{equation}
$\vec{\lambda}=(\lambda_1, \cdots,\lambda_M, \lambda'_1,\cdots, \lambda'_M)$ and $\vec{z}=(u_1,\cdots,u_M,v_1,\cdots,v_M)$. 

It is required that $\beta N \Omega\to\infty$ for the existence of SPT, because if the thermal energy $1/\beta$ is of the same order of $N\Omega$, the system will always be in SP \cite{Peng2019}. Therefore, we can use Laplace's method \cite{Wang1973} to approximate $Z$ Eq. \eqref{z1} as
\begin{equation}
Z\approx (\frac{2}{\beta})^{M}|h(\vec{z}_{\mathrm{min}})|^{-\frac{1}{2}}(\prod_{i=1}^{M}{\omega_{i}^{-1}})e^{-\beta N\Omega\phi(\vec{z}_{\mathrm{min}})},
\label{tz}
\end{equation}
where $|h(\vec{z}_{\mathrm{min}})|$ is the Hessian determinant of $\phi(\vec{z})$ at the global minimum point $\vec{z}_{\mathrm{min}}$. Therefore, the rescaled free energy $f=F/N\Omega=-lnZ/N\Omega\beta$ is well approximated by the global minimum of $\phi(\vec{z})$. So  $\phi(\vec{z})$ acts as a Landau potential.
 
The rescaled mean photon number $\langle a^{\dag}_i a_i\rangle/\eta_i$ is given by
\begin{eqnarray}
\frac{\langle a^{\dagger}_i a_i\rangle}{\eta_{i}}&=&\frac{\text{Tr}[a^{\dagger}_i a_i\exp(-\beta H)]} {\eta_i\text{Tr}[\exp(-\beta H)]}=-\frac{1}{\eta_i}\cdot\frac{\partial \ln{\tilde{Z}}}{\partial(\beta\omega_i)}\nonumber\\
&=&\frac{\int^{+\infty}_{-\infty}{(u^{2}_i + v^{2}_i)\mathrm{exp}[-\beta N\Omega\phi(\vec{z})]d\vec{z}}}{\int^{+\infty}_{-\infty}{\mathrm{exp}[-\beta N\Omega\phi(\vec{z})]d\vec{z}}}.
\label{rescalp}
\end{eqnarray}
By Laplace's method, it immediately follows that $\frac{\langle a^{\dagger}_i a_i\rangle}{\eta_{i}}=u^2_{i\min}+v^2_{i\min}$, and $\sum \frac{\langle a^{\dagger}_i a_i\rangle}{\eta_{i}}=|\vec{z}_{\min}|^2$, where $u_{i\min}$, $v_{i\min}$ and  $|\vec{z}_{\min}|$ are the global minima of the Landau potential $\phi(\vec{z})$. 
So $\langle a^{\dag}_i a_i\rangle/\eta_i$ acts as the order parameter for the $i$-th mode, which is the location of the GMLP, being zero for NP and finite for SP.

Therefore, we propose a concise method to study SPT just by determining whether the origin of a Landau potential is the global minimum. Accordingly, we obtain
a general phase diagram feature by studying the Landau potential $\phi(\vec{z})$ Eq.~\eqref{landaup}, which gives
\begin{eqnarray}
z_i\cdot\frac{\partial\phi}{\partial z_i}&=&2z^{2}_i+\sum_{k}{(\lambda_k\cdot z_i)\frac{\partial\mu}{\partial(\lambda_k\cdot z_i)}},
\label{eq:three}
\\
\lambda_k\cdot\frac{\partial\phi}{\partial\lambda_k}&=&\sum_{i}{(\lambda_k\cdot z_i)\frac{\partial\mu}{\partial(\lambda_k\cdot z_i)}}.
\label{eq:four}
\end{eqnarray}
Combining Eqs.~\eqref{eq:three} and ~\eqref{eq:four}, we finally obtain 
\begin{equation}
\vec{z}\cdot\triangledown_{\vec{z}}\phi(\vec{\lambda},\vec{z})=2|\vec{z}|^2+\vec{\lambda}\cdot\triangledown_{\vec{\lambda}}\phi(\vec{\lambda},\vec{z}).
\label{eq:five}
\end{equation}   
Since the Landau potential is continuous and differentiable with respect to the coherent state parameters $\vec{z}$, the GMLP must be an extremum, meaning that the left hand side of Eq.~\eqref{eq:five} equals zero at the minimum point $\vec{z}_{\min}$. Since $\vert \vec{z}_{\min}\vert^2$ is zero for NP and positive for SP, $\vec{\lambda}\cdot\triangledown_{\vec{\lambda}}\phi(\vec{\lambda},\vec{z}_{\min})$ is zero and negative for NP and SP respectively. This means that increasing $\vec \lambda$ radially will not change $\phi(\vec{\lambda},0)$. However, if there is an extreme other than zero, then $\phi(\vec{\lambda},\vec{z}_{\min})$ will decrease radially. As can be seen in Eq.~\eqref{landaup}, $\phi(\vec{z}=0)$ is the GMLP when $\vert\vec{\lambda}\vert\approx0$. If there is a nonzero solution to $\phi(\vec{\lambda}_c,\vec{z}\neq 0)=\phi(\vec{\lambda}_c,0)$ as $\vec{\lambda}$ increases, then $\phi(\vec{\lambda},\vec{z}\neq0)$ will always be less than $\phi(\vec{\lambda},0)$ in the radial direction of $\vec{\lambda}$. To conclude, the origin is in NP in the coupling parameter $\vec{\lambda}$ space. It goes to SP radially and will never go back. The phase boundary can be obtained by solving $\phi(\vec{\lambda}_c, \vec{z}_c)=\phi(\vec{\lambda}_c, 0)$, where $\vec{z}_c$ is an extreme point. If $\vec{z}_c=0$, then the  order parameter rescaled mean photon number changes continuously from $0$ to nonzero and the SPT is $2nd$-order. If $\vec{z}_c\neq0$, then the SPT is $1st$-order. This can be understood as $|\vec{z}_{\min}|^2$ is proportional to the first derivative of the free energy as $|\vec{z}_{\min}|^2=-\vec{\lambda}\cdot\triangledown_{\vec{\lambda}}\phi(\vec{\lambda},\vec{z}_{\min})$, according to Eq. \eqref{eq:five}.
If there is no nonzero solution $\vec{z}_c$ at any nonzero $\vert\vec{\lambda}\vert$, then NP only exists at $\vert\vec{\lambda}\vert=0$ and there is no SPT. Because the GMLP is clearly not located at the origin when $|\vec{\lambda}|\to\infty$, and it moves to nonzero values once  $|\vec{\lambda}|\neq0$. We will illustrate these results with the multimode Dicke model with both one- and two-photon terms, the Rabi-Stark-Hubbard model, and the anisotropic Rabi-Stark model.

\section{\label{sec:level3}Multimode Dicke model with both one- and two-photon terms}

The general structure of the phase diagram of single-mode Dicke model with both one- and two-photon terms has been discussed in Ref.~\citep{Peng2019}, which agrees with our analysis. However, it remains unknown whether the same phase diagram feature holds for the multimode case. To answer this question, we extend it to the multimode case and obtain the Hamiltonian by only keeping the 1$st$, 2$nd$, 4$th$ and 5$th$ terms with $\alpha_{ij}=\beta_{ij}=1$  in Eq.~\eqref{eq:one}. The Hamiltonian for the model reads ($\hbar=1$)
\begin{eqnarray}
H&=&\sum_{\nu=1}^{M}{\sum_{i=1}^{N}{\left\{\frac{g_{\nu}}{\sqrt{N}}(a_{\nu}+a^{\dag}_{\nu})+
{\frac{g^{\prime}_{\nu}}{N}[a^{2}_{\nu}+(a^{\dag}_{\nu})^2]}\right\}\sigma^{x}_{i}}}\nonumber\\
& &+
\sum_{\nu=1}^{M}{\omega_{\nu} a^{\dag}_{\nu}a_{\nu}}+\sum_{i=1}^{N}{\frac{\Omega}{2}\sigma^{z}_{i}},
\label{eq:six}
\end{eqnarray}
where $a_{\nu}$ ($a^{\dag}_{\nu}$) denotes the annihilation (creation) operator with frequency $\omega_{\nu}$ and $\sigma^{x,z}_{i}$ are Pauli operators with transition frequency $\Omega$. Here, $N$ and $M$ are the number of qubits and finite bosonic modes, respectively. Moreover, $g_{\nu}$ and $g^{\prime}_{\nu}$ are the cavity-qubit coupling constants in the one-photon and two-photon processes for the corresponding bosonic modes, respectively. The stability condition of the system is $g^{\prime}_{\nu}<\omega_{\nu}/2$ [see the Appendix]. Note that the coexistence of both one- and two-photon terms leads to the vanishing of both $Z_2$ symmetry in the Dicke model and $Z_4$ symmetry in the two-photon Dicke model. In general, it will be $1st$ order if a phase transition exists in a system without any symmetry.  

First, to consider the combined limit $\eta_{\nu}=N\Omega/\omega_{\nu}\rightarrow\infty$, we change the Hamiltonian into
\begin{eqnarray}
H&=&\Omega\sum_{\nu=1}^{M}{\sum_{i=1}^{N}{\left\{\frac{\gamma_{\nu}}{2}\frac{(a_{\nu}+a^{\dag}_{\nu})}{\sqrt{\eta_{\nu}}}+
{\frac{\gamma^{\prime}_{\nu}}{2}\frac{a^{2}_{\nu}+(a^{\dag}_{\nu})^2}{\eta_{\nu}}}\right\}\sigma^{x}_{i}}}\nonumber\\
& &+
\sum_{\nu=1}^{M}{N\Omega\frac{a^{\dag}_{\nu}a_{\nu}}{\eta_{\nu}}}+\sum_{i=1}^{N}{\frac{\Omega}{2}\sigma^{z}_{i}},
\label{eq:seven}
\end{eqnarray}
where $\gamma_{\nu}=2g_{\nu}/\sqrt{\Omega\omega_{\nu}}$ and $\gamma^{\prime}_{\nu}=2g^{\prime}_{\nu}/\omega_{\nu}$. 
According to the analysis in the last section, we apply the mean-field method proposed in Ref.~\citep{Wang1973} to obtain the partition function 
\begin{eqnarray}\label{z1d}
Z = \left( \prod_{\nu=1}^{M} \eta_\nu \right) \pi^{-M} \int e^{-\beta N \Omega |\vec{z}|^2}\rm{Tr}_qe^{-\beta\sum_{i=1}^N h_i} d\vec{z}, 
\end{eqnarray}
where 
\begin{equation}\label{hima}
h_i=\frac{\Omega}{2}[\sigma_i^z+\xi\sigma_i^x]
\end{equation}
with
\begin{eqnarray}
\xi&=&2\sum_{\nu=1}^{M}{\left[\gamma_{\nu}u_{\nu}+\gamma^{\prime}_{\nu}(u^{2}_{\nu}-v^{2}_{\nu})\right]},\\
\vec{z}&=&(u_{1},u_{2},\cdots,u_{M},v_{1},v_{2},\cdots,v_{M}).    
\end{eqnarray}
Since $h_i$ is independent of each other, Eq. \eqref{z1d} reduces to
\begin{eqnarray}\label{z2d}
Z = \left( \prod_{\nu=1}^{M} \eta_\nu \right) \pi^{-M} \int e^{-\beta N \Omega |\vec{z}|^2}(\rm{Tr}_qe^{-\beta h_i} )^Nd\vec{z}.
\end{eqnarray}
It is easy to find $\rm{Tr}_qe^{-\beta h_i}=2Cosh[\beta\Omega\sqrt{1+\xi^2}/2]$, so we finally obtain the Landau potential 
\begin{equation}
\phi(\vec{z})=\sum_{\nu=1}^{M}{(u^2_{\nu}+v^2_{\nu})}
-\frac{1}{\beta\Omega}\ln{\left[2\cosh\left({\frac{\beta\Omega}{2}\sqrt{1+\xi^2}}\right)\right]}
\label{eq:nine}
\end{equation}
with the partition function takes the form of Eq. \eqref{z1}. 

\subsection{Zero-temperature case}\label{zerot}
For simplicity, we first focus on the zero temperature case $\beta\rightarrow\infty$ where the system is in the ground state. To obtain the GMLP, we first analyze the extremal behaviors of the Landau potential 
\begin{eqnarray}
\frac{\partial\phi}{\partial u_{m}}&=&2u_{m}-(\gamma_{m}+2\gamma^{\prime}_{m}u_m)\frac{\xi}{\sqrt{1+\xi^2}}=0,
\label{eq:thirteen}
\\
\frac{\partial\phi}{\partial v_{m}}&=&2v_{m}(1+\gamma_{m}^{\prime}\frac{\xi}{\sqrt{1+\xi^2}})=0. 
\label{eq:fourteen}
\end{eqnarray}
When the coupling strength is vanishingly small, the system stays in the NP since the GMLP is located at the origin $\vec{z}=0$. As the coupling strength $\vec{\lambda}$ increases, the position of GMLP may shifts from the origin to a non-origin point. The presence of the SP requires certain nonzero solutions to Eqs.~\eqref{eq:thirteen} and \eqref{eq:fourteen}. Noting $|\xi/\sqrt{1+\xi^2}|<1$, the GMLP is always located at the origin in the $v_{\nu}$ plane under the stability condition of $\gamma^{\prime}_{\nu}<1$. Neglecting the $v$ part, we rewrite the landau potential as
\begin{equation}
\phi(\vec{u})=\sum_{\nu=1}^{M}u_{\nu}^{2}-\frac{1}{2}\sqrt{1+4\left[\sum_{\nu=1}^{M}{(\gamma_{\nu}u_{\nu}+\gamma_{\nu}^{\prime}u_{\nu}^{2})}\right]^{2}},
\label{eq:fifteen}
\end{equation}
with $\vec{u}=(u_{1},u_{2},\cdots,u_{M})$. Accordingly, we rewrite Eq.~\eqref{eq:thirteen} as 
\begin{equation}
\frac{\partial\phi}{\partial u_{m}}=2u_{m}-\frac{2\sum_{\nu=1}^{M}{(\gamma_{\nu}u_{\nu}+\gamma_{\nu}^{\prime}u_{\nu}^{2})}(\gamma_{m}+2\gamma_{m}^{\prime}u_{m})}{\sqrt{1+4\left[\sum_{\nu=1}^{M}{(\gamma_{\nu}u_{\nu}+\gamma_{\nu}^{\prime}u_{\nu}^{2})}\right]^{2}}}=0.
\label{eq:sixteen}
\end{equation}
Clearly, $u_{\nu}=0$ is a trivial solution to Eq.~\eqref{eq:sixteen}. According to Eq.~\eqref{eq:sixteen}, we take $u_1$ as the unique variable and then $u_\nu$ satisfies 
\begin{equation}
u_{\nu}=\frac{\gamma_{\nu}u_{1}}{\gamma_1+2(\gamma_1^{\prime}-\gamma_{\nu}^{\prime})u_{1}}.
\label{eq:seventeen}
\end{equation} 
Substituting Eq.~\eqref{eq:seventeen} into Eq.~\eqref{eq:fifteen}, the Landau potential turns to be a one-variable function with respect to $u_{1}$. However, it is rather complicated to determine the global minimum by only taking the derivatives of the Landau potential like ordinary cases \cite{Wang1973}, because it is difficult to determine when Eq. \eqref{eq:sixteen} has nonzero solutions $u_1$ even for $\gamma_{\nu}^{\prime}=\gamma^{\prime}$, let alone judging wether $\phi(u_1)<\phi(0)$.
Nonetheless, we can determine the critical condition analytically by using our concise method of solving $\phi(u_{1}\neq 0)=\phi(0)$ without need of taking derivatives,
which reduces to a simple quadratic equation when  $\gamma_{\nu}^{\prime}=\gamma^{\prime}$
\begin{equation}
(1-\gamma^{\prime 2})\gamma^2 u^{2}_1-2\gamma^{\prime}\gamma_1\gamma^2 u_1+(1-\gamma^2)\gamma^2_1=0,
\label{eq:eighteen}
\end{equation}
where $\gamma^{2}=\sum_{\nu}{\gamma_{\nu}^{2}}$. The existence condition of nonzero solutions to Eq.~\eqref{eq:eighteen} leads to a phase boundary $\gamma^{2}+\gamma^{\prime 2}=1$, which reduces to that in the multimode QRM by choosing $\gamma^{\prime}=0$ \citep{Shen2021, Peng2019ar} or in the single-mode QRM with both one- and two-photon terms by replacing $\sum _\nu \gamma_\nu^2$ with a single $\gamma_\nu^2$ \citep{Peng2019, Ying2020}. Interestingly, this means that the coupling strength requirement on SPT for the single mode case is much reduced for the multimode case, even possibly to the strong coupling regime. This multimode cooperative behavior could make SPT easier. Furthermore, according to Eq.~\eqref{eq:sixteen} and \eqref{eq:seventeen}, we can readily obtain the analytical expression of the rescaled mean photon number $u^2_{\nu}=(1-\gamma^{-4})\gamma^2_{\nu}/4$ with a phase boundary $\gamma^2=1$ for the multimode QRM by taking $\gamma^{\prime}_{\nu}=0$. The expression indicates that the SPT is $2nd$ order.
\begin{figure}[t]
\centering
\renewcommand\figurename{\hspace{1 em} FIG.}
\subfigure{\includegraphics[width=0.48\linewidth, height= 1.1 in]{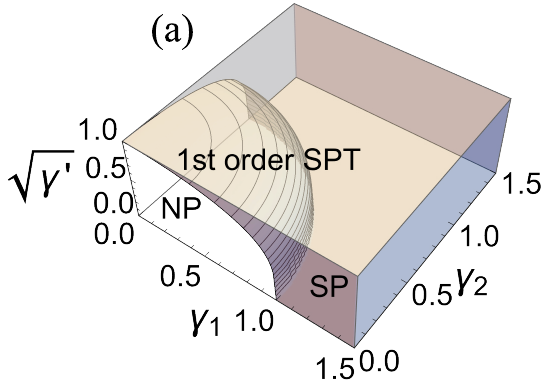}\label{fig:1(a)}}\hspace{-1 mm}
\subfigure{\includegraphics[width=0.48\linewidth, height= 1.1 in]{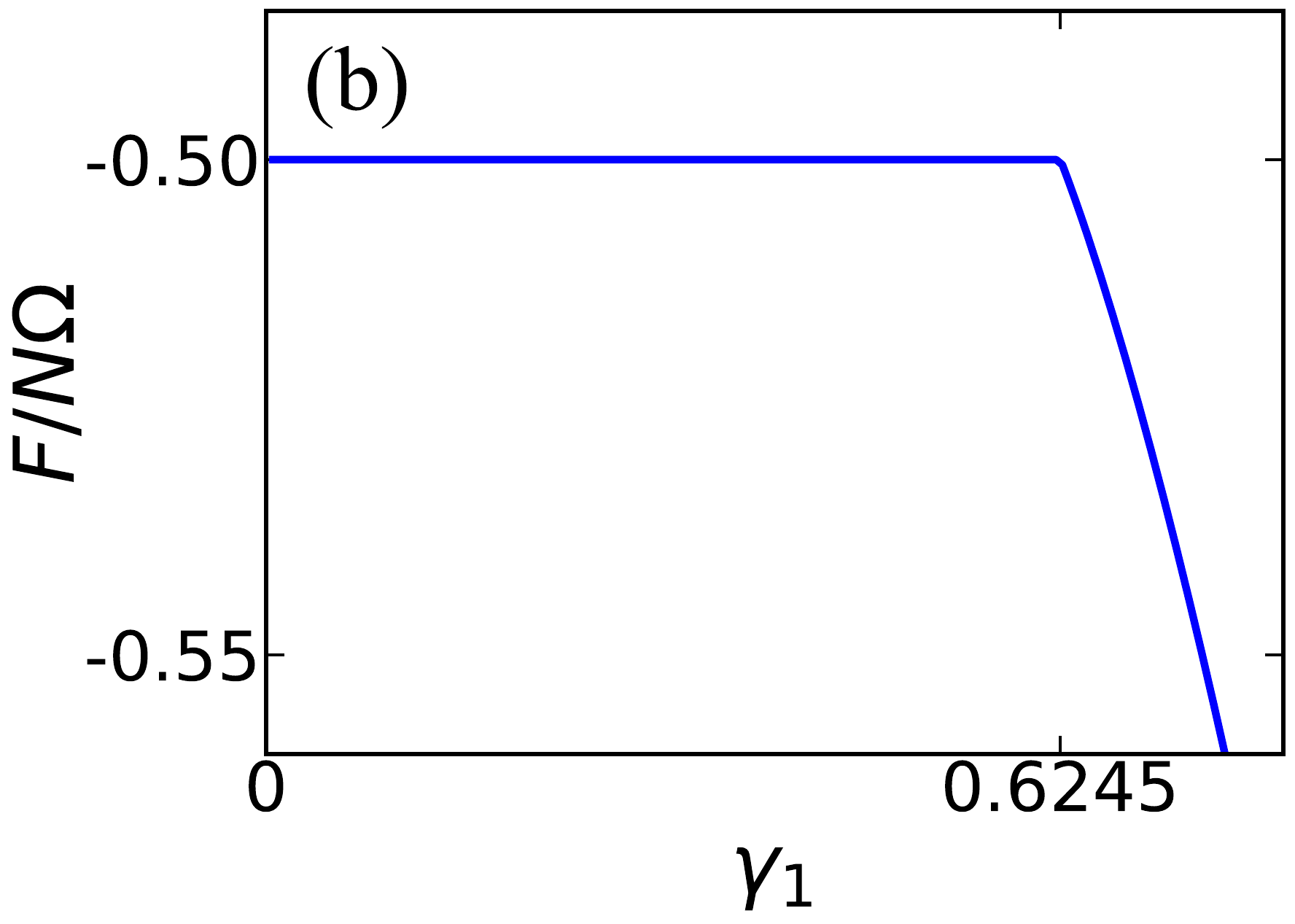}\label{fig:1(b)}}\vskip -8 pt
\subfigure{\includegraphics[width=0.48\linewidth, height= 1.1 in]{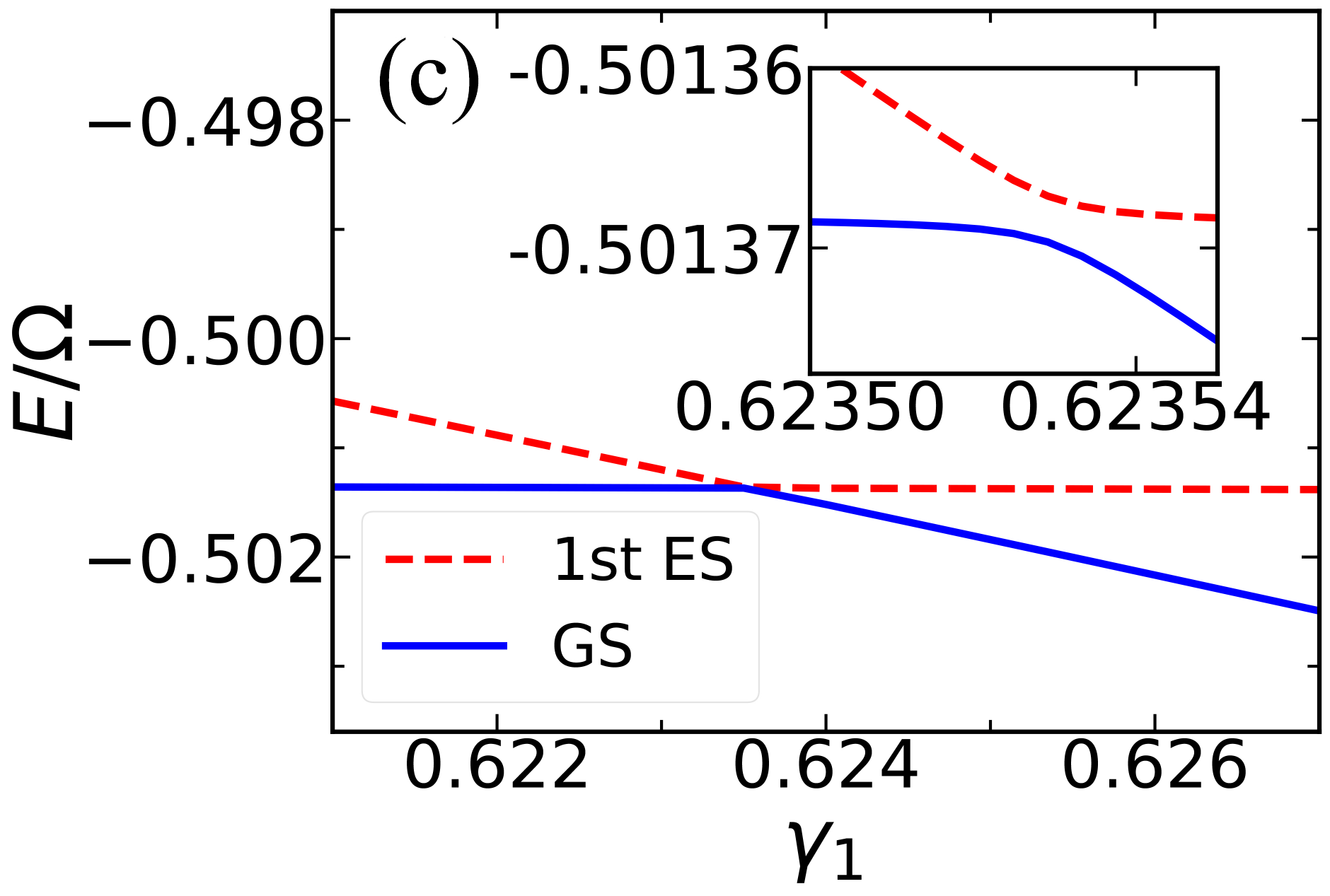}\label{fig:1(c)}}\hspace{-1 mm}
\subfigure{\includegraphics[width=0.48\linewidth, height= 1.1 in]{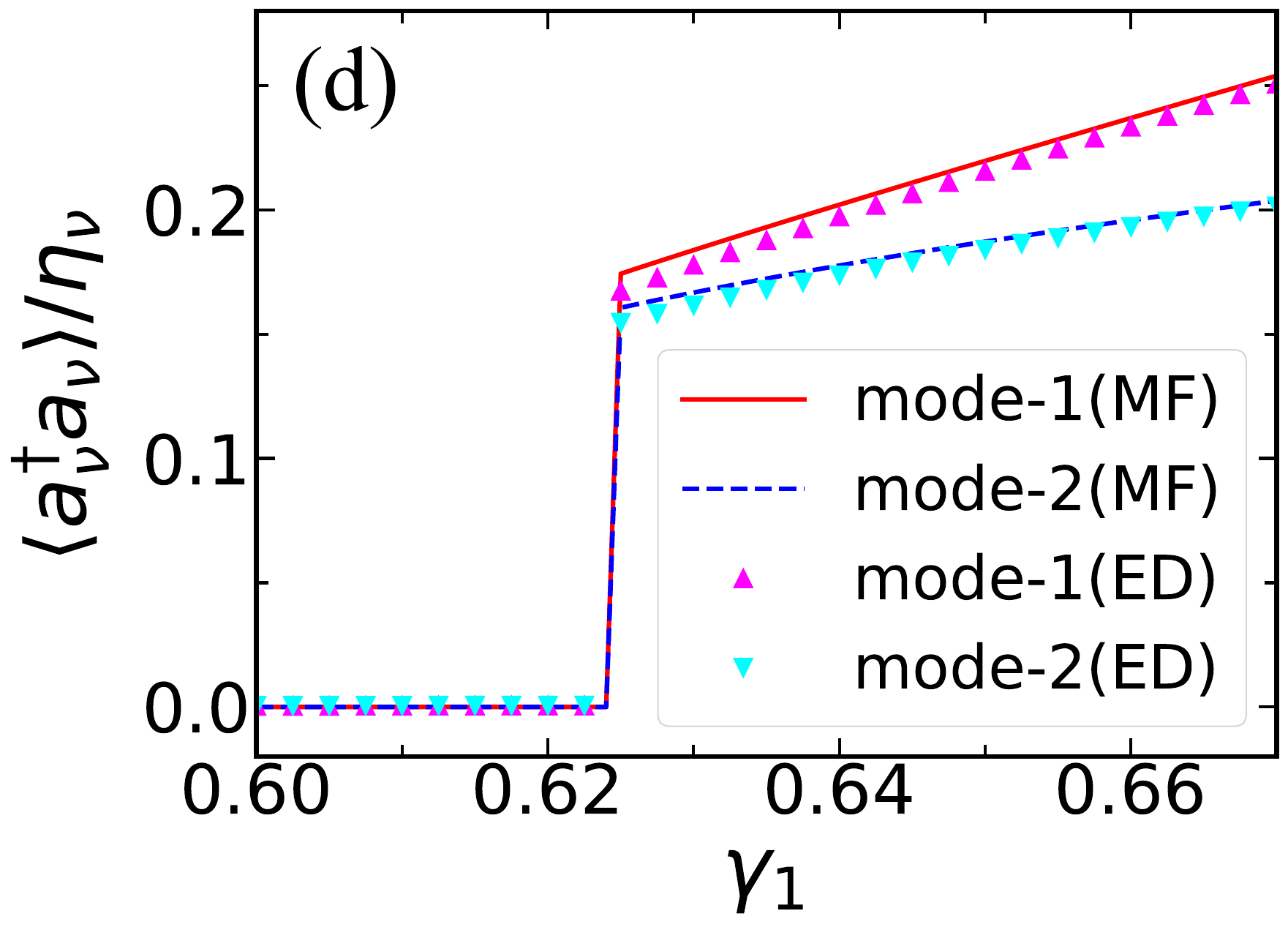}\label{fig:1(d)}}
\caption{\justifying\label{fig:1}(a) The phase diagram of two-mode Dicke model with both one- and two-photon terms depicted from the phase boundary at $T=0$ in the thermodynamic limit. (b) The reduced free energy as the function of $\gamma_1$ plotted numerically in the classical oscillator limit for finite $N$ where the critical value $\gamma_{1c}=0.6245$, $\gamma_2=0.6$, and $\gamma^{\prime}_{1}=\gamma^{\prime}_{2}=0.5$. (c) The ground state (GS) and the first excited state (1st ES) energies as a function of $\gamma_1$ numerically obtained by exact diagonalization (ED) for $N=1$ at the finite frequency ratios $\eta_1=200$, $\eta_2=180$. The inset shows an avoided level-crossing in the neighbourhood of the critical point. (d) The rescaled mean photon number of the GS for two modes numerically obtained from mean-field method (MF) and ED, respectively.}
\end{figure} 
We also obtain the solution $u_{1c}=\gamma_1\gamma^{\prime}/(1-\gamma^{\prime 2})$ to Eq.~\eqref{eq:eighteen} implying $\vert \vec{z}_c\vert\neq0$ at the boundary $\gamma^{2}+\gamma^{\prime 2}=1$, no matter how vanishingly small $\gamma^{\prime}$ is, as long as $\gamma^{\prime}\neq 0$. It means that the rescaled mean photon number (first derivative of the reduced free energy) is discontinuous at the phase transition point, and the SPT is $1st$ order. However, when $\gamma^{\prime}$ increases smoothly from zero to nonzero, the rescaled mean photon number undergoes a continuous change and the SPT is $2nd$ order. 
   
We draw the phase diagram of two-mode QRM with both one- and two-photon terms to show the general phase diagram feature that a SPT radially occurs along $\vec{\lambda}=(\gamma_1, \gamma_2, \sqrt{\gamma^{\prime}})$ under the stability condition in Fig.~\ref{fig:1(a)}. Figure \ref{fig:1(b)} shows the free energy per qubit in the unit of $\Omega$, plotted by numerically finding the global minima of the Landau potential Eq.~\eqref{eq:fifteen} with respect to $\gamma_1$. The discontinuity of the first derivative of free energy implies a $1st$ order phase transition, testified by the exact diagonalization (ED) at finite frequency ratios $\eta_1=200$, $\eta_2=180$, as shown in Fig.~\ref{fig:1(c)}, where the sharp avoided crossing brings discontinuity when $\eta_{1,2}\to\infty$. Meanwhile, the cavity field in the ground state acquires a macroscopic population of photons [see Fig.~\ref{fig:1(d)}]. The result of ED coincides with that obtained by finding the minimum of Landau potential Eq.~\eqref{eq:fifteen}. We also numerically calculate the ratio of the rescaled mean photon number between two modes in the SP and find it consistent with Eq.~\eqref{eq:seventeen}. Here, the Fock subspace of each field mode is truncated at $N_{\mathrm{cut}}=100$ to realize the convergence of the ground state. 

\subsection{Finite-temperature case}{\label{sec:ft}}
The temperature dependence of the critical condition is important in
experiments, so we study it in this subsection. All we need to do is to determine the GMLP of Eq. \eqref{eq:nine}. We first analyze the extremal behaviors of the Landau potential 
\begin{eqnarray}
\frac{\partial\phi}{\partial u_{m}}&=&2u_{m}-(\gamma_{m}+2\gamma^{\prime}_{m}u_m)\nonumber\\
&&\times\frac{\tanh(\beta\Omega\sqrt{1+\xi^2}/2)\xi}{\sqrt{1+\xi^2}}=0,
\label{lant1}
\\
\frac{\partial\phi}{\partial v_{m}}&=&2v_{m}\left[1+\gamma_{m}^{\prime}\frac{\tanh(\beta\Omega\sqrt{1+\xi^2}/2)\xi}{\sqrt{1+\xi^2}}\right]=0. 
\label{lant2}
\end{eqnarray}
Obviously, Eq. \eqref{lant2} has no nonzero solutions since $\gamma_{m}^{\prime}\frac{\tanh(\beta\Omega\sqrt{1+\xi^2}/2)\xi}{\sqrt{1+\xi^2}}<1$. So $v_{m\min}$ always equals zero, and we only need to find $u_{m\min}$. Substituting $v_m=0$ into Eq. \eqref{lant1}, we find Eq. \eqref{eq:seventeen} is still valid, which can be used to simplify a set of differential equations on $u_m$ to just $u_1$. However, we can not find the phase transition points by determining when Eq. \eqref{lant1} has nonzero solutions. So we combine our concise method of
\begin{equation}
\phi(\vec{z}_c, \beta_c)= \phi(0, \beta_c),
\end{equation}
and Eq. \eqref{lant1} to find the critical $\vec{z}_c$ and $\beta_c$  for certain system coupling parameters.

\begin{figure}[!htb]
\renewcommand\figurename{\hspace{1 em} FIG.}
\includegraphics[width=0.91\linewidth]{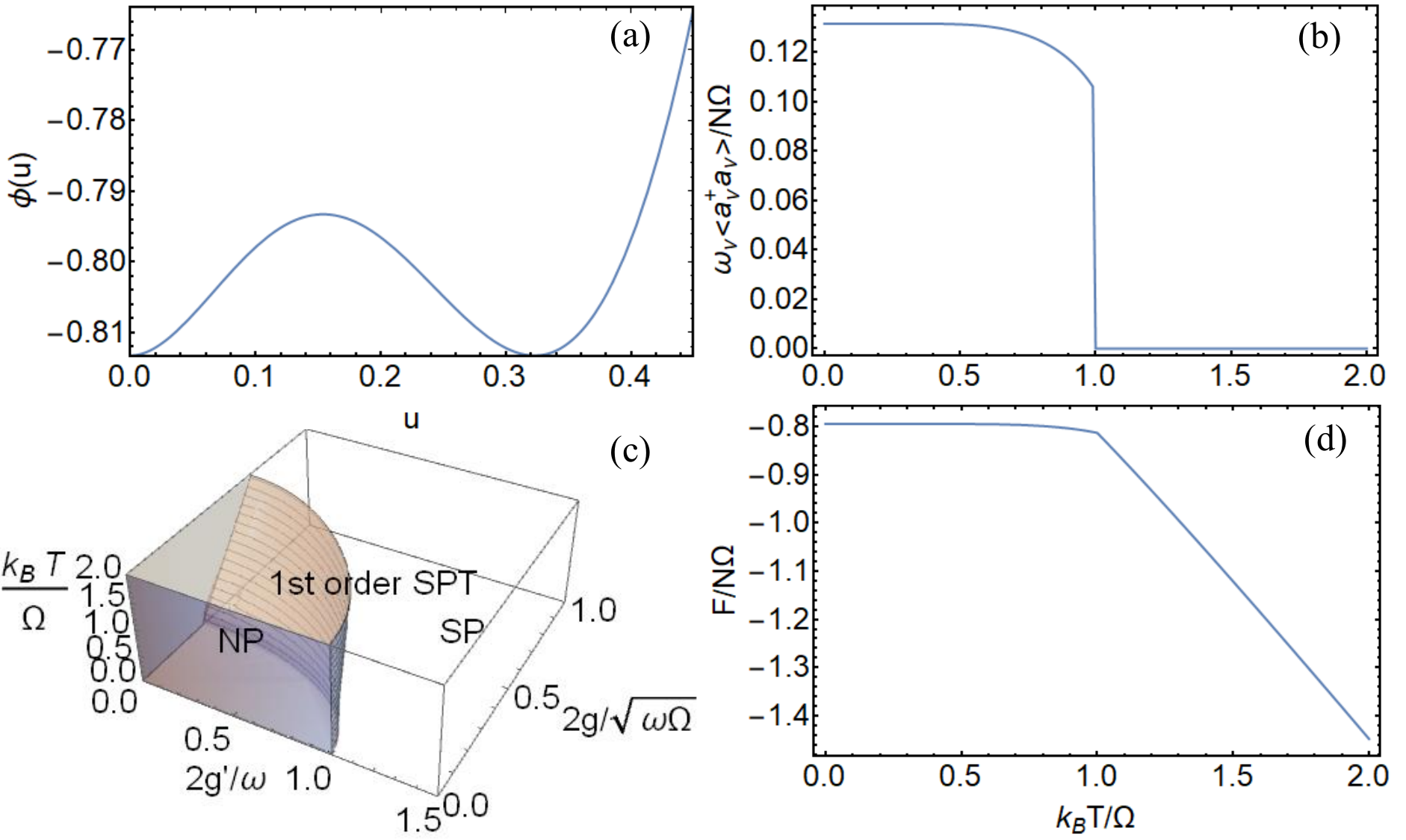}
\caption{\justifying \label{figad}SPT properties of $H$ Eq. \eqref{eq:seven} for $M=10$, $N\to\infty$, $g_\nu=g$, $g'_\nu=g'$,$\omega_\nu=\omega$. (a) Landau potential at the phase transition point.  $k_B T_c/\Omega=1$, $2g/\sqrt{\omega\Omega}=0.3839$,  $2g'/\omega=0.5$ and $u_c=0.323$. (b) Rescaled mean photon number as a function of temperature. $2g/\sqrt{\omega\Omega}=0.3839$,  $2g'/\omega=0.5$. (c) Phase diagram. (d) Rescaled free energy as a function of temperature. $2g/\sqrt{\omega\Omega}=0.3839$,  $2g'/\omega=0.5$.}
\end{figure} 
We illustrate this method by choosing $M=10$, $\omega_\nu=\omega$, $g_\nu=g$, $g'_\nu=g'$ and $N\to\infty$. Note that SPT is only dependent on temperature under the thermodynamic limit $N\to\infty$ \cite{Ashhab2013,Peng2019}, because $N\beta\Omega\to\infty$ is required for SPT. If $N$ is finite and $\beta\Omega\to\infty$, then $\beta\Omega$ is canceled out in the Landau potential. Physically, the thermal energy is vanishing small even compared to a single qubit energy. We draw the Landau potential at the critical $k_B T_c/\Omega=1$, $2g/\sqrt{\omega\Omega}=0.3839$,  $2g'/\omega=0.5$ and $u_c=0.323$ in Fig. \ref{figad} (a). There is an nonzero extreme $u_c$ with the same height as the origin, signifying the occurring of a phase transition at this point. We further draw the rescaled  mean photon number and free energy with respect to temperature in Fig. \ref{figad} (b) and (d). The discontinuity of the first-order derivative of free energy indicates a first-order SPT. A phase diagram with respect to $g$, $g'$ and $k_B T$ is shown in Fig. \ref{figad} (c). As mode number increases, the phase transition point for couplings $g$ decreases as $2g/\sqrt{\omega\Omega}=\sqrt{(1-4g’^2/\omega^2)/M}$ at zero temperature.

\subsection{Disorder}
Another concern is that the disorder could influence the
SPT. Recent works by H.-T. Chen and A. Nitzan \textit{et al.} \cite{PhysRevA.106.053703} and S. F. Yelin \textit{et al.} \cite{PhysRevA.109.013720} discuss various disorder effects on creating subradiant
states that diminish superradiance. Here we study how disorder
affects the overall phase diagram by introducing the disorder-induced inhomogeneities on the qubit frequencies \cite{PhysRevA.106.053703} and coupling strengths \cite{PhysRevA.109.013720} to the typical multimode Dicke model
\begin{equation}
H=\sum_{\nu}\omega_\nu a_\nu^\dagger a_\nu+\sum_{i=1}^N \sum_{\nu=1}^M\frac{g_{i\nu}}{\sqrt{N}}(a_\nu+a_\nu^\dagger)\sigma_{ix}+\sum_{i=1}^N\frac{\Omega_i}{2}\sigma_{iz}.
\end{equation}
Expanding $Z=\rm{Tr}e^{-\beta H}$ in coherent state basis, we obtain the same formula as Eq. \eqref{z1d} with
\begin{eqnarray}
h_i&=&\frac{\Omega_i}{2}\sigma_i^z+\xi_i\sigma_i^x,\\
\xi_i&=&2\sum_{\nu=1}^{M}{\left[\gamma_{ i\nu}u_{\nu}\right]},\\
\vec{z}&=&(u_{1},u_{2},\cdots,u_{M},v_{1},v_{2},\cdots,v_{M}),\\
u_\nu&=&x_\nu\sqrt{\omega_\nu}/\sqrt{N\Omega},v_\nu=y_\nu\sqrt{\omega_\nu}/\sqrt{N\Omega},
\end{eqnarray}
where $\Omega$ is a positive generalized mean of qubit energies $\{\Omega_i \}$.
Tracing out the qubit part, we obtain the Landau potential 
\begin{equation}
\phi(\vec{z})=\sum_{\nu=1}^{M}{(u^2_{\nu}+v^2_{\nu})}
-\frac{1}{\beta N\Omega}\sum_{i=1}^N\ln{\left[2\cosh\left({\frac{\beta\Omega}{2}\sqrt{  \delta_i+\xi_i^2}}\right)\right]},   
\end{equation}
where $\delta_i=\Omega_i/\Omega$. To find out its global minimun, we study the extremal behaviors 
\begin{eqnarray}
\frac{\partial\phi}{\partial u_{m}}=&&2u_{m}-\sum_{i=1}^N\tanh({\frac{\beta\Omega\sqrt{\delta_i+\xi_i^2}}{2}})\nonumber\\&&\times\frac{\gamma_{m}\xi_i}{N\sqrt{\delta_i+\xi_i^2}}=0,
\label{disorder1}
\\
\frac{\partial\phi}{\partial v_{m}}=&&2v_{m}=0. \label{disorder2}
\end{eqnarray}
Obviously, the global minimum of $v_m$ is at the origin. However, it is difficult to find the critical condition for Eq. \eqref{disorder1} to have nonzero solutions. As mentioned in Sec. \ref{sec:ft}, we can obtain the critical condition numerically by combining $\phi(\vec{z}_c)=\phi(0)$ and Eq. \eqref{disorder1}.

We study the SPT properties analytically in two specific cases. First, we assume $g_{i\nu}=g_\nu$ as in \cite{PhysRevA.106.053703}, so that $\zeta_i=\zeta$, $\gamma_{im}=\gamma_m$. We can obtain
$u_m=u_1\gamma_m/\gamma_1$ from Eq. \eqref{disorder1}, and reduce it to 
\begin{equation}
u_1-\frac{1}{N}\sum_{i=1}^N \tanh[\frac{\beta\Omega}{2}\sqrt{\delta_i^2+\xi^2}]\frac{\gamma^2u_1}{\sqrt{\delta_i^2+\xi^2}}=0,
\end{equation}
where $\gamma^2=\sum_{\nu=1}^M\gamma_{\nu}^2$. So that 
\begin{equation}\label{l5}
\frac{1}{\gamma^2}=\frac{1}{N}\sum_{i=1}^N \tanh[\frac{\beta\Omega}{2}\sqrt{\delta_i^2+\xi^2}]\frac{1}{\sqrt{\delta_i^2+\xi^2}}.
\end{equation}
According to similar analysis in \cite{Wang1973}, solution to Eq. \eqref{l5} exists when
\begin{equation}\label{l6}
\frac{1}{\gamma^2}<\frac{1}{N}\sum_{i=1}^N \tanh[\beta\Omega\delta_i/2]{\frac{1}{\delta_i}}=\frac{1}{\gamma_c^2}.
\end{equation}
Meanwhile, if we know priorly the SPT of the Dicke model is second-order, we can substitute $u_c=0$ into Eq. \eqref{l5} to obtain $\gamma_c$ directly.

Now we analyze the disorder effect. According to \cite{PhysRevA.106.053703}, we assume a random
 component in the qubit transition frequency:
\begin{equation}
\Omega_i=\Omega+\delta\Omega_i
\end{equation}
with $\langle\delta\Omega_i\rangle=0$. We can determine the phase boundary by substituting the distribution $\Omega_i$ into Eq. \eqref{l6}. Here we do a brief study on zero temperature case ($\beta\to\infty$), where
\begin{equation}
\frac{1}{\gamma^2}<\frac{1}{N}\sum_{i=1}^N {\frac{1}{\delta_i}}=\frac{1}{\gamma_c^2}.
\end{equation}If all the qubits are identical, then we recover the SPT condition $
\gamma_c^2=1$ in Sec. \ref{zerot}. If $\langle\delta\Omega_i\rangle=0$, then $
\gamma_c^2<1$, so that the phase boundary is shifted.

Next we consider the single-mode Dicke model, whose Landau potential reads
\begin{equation}
\phi(\vec{z})=u^2+v^2
-\frac{1}{\beta N\Omega}\sum_{i=1}^N\ln{\left[2\cosh\left({\frac{\beta\Omega}{2}\sqrt{  \delta_i+4\gamma_i^2 u^2}}\right)\right]},
\end{equation}
where $\gamma_i=g_i\sqrt{\omega}/\sqrt{N\Omega}$.
The extremes satisfy
\begin{eqnarray}
\frac{\partial\phi}{\partial u}=&&2u-\sum_{i=1}^N\tanh({\frac{\beta\Omega\sqrt{\delta_i+4\gamma_i^2u^2}}{2}})\nonumber\\&&\times\frac{2\gamma_i^2 u}{\sqrt{\delta_i+4\gamma_i^2u^2}}=0,
\label{disorder1a}
\\
\frac{\partial\phi}{\partial v}=&&2v=0. \label{disorder2a}
\end{eqnarray}
According to similar analysis, SP exists if 
\begin{equation}\label{nbo}
\frac{1}{N}\sum_{i=1}^N \tanh[\beta\Omega\delta_i/2]{\frac{\gamma_i^2}{\delta_i}}<1.
\end{equation}
Substituting inhomogeneous $\gamma_i$ and $\delta_i$ induced by disorder into Eq. \eqref{nbo}, we can obtain the new phase boundary.  
To conclude, disorder clearly shift the phase boundary, which can be determined by above studies.

\section{\label{sec:level4}Rabi-Stark-Hubbard model}

Besides the one- and two-photon coupling processes, we also study the effects of the nonlinear Stark coupling and inter-cavity hopping on the phase diagram. To this end, we consider one-dimensional Rabi-Stark-Hubbard model, describing a chain of cascaded cavities with $M$ sites where each cavity is labeled with index $i$ and governed by Rabi-Stark model. The adjacent cavities are coupled though photon couplings. This Hamiltonian corresponds to Eq.~\eqref{eq:one} without the third and fifth lines, which reads
\begin{equation}
H=\sum^{M}_{j}{H^{\mathrm{RS}}_{j}}-J\sum^{M}_{j}{(a_{j}+a^{\dag}_{j})(a_{j+1}+a^{\dag}_{j+1})},
\label{eq:nineteen}
\end{equation}
\begin{figure}[!htb]
\renewcommand\figurename{\hspace{1 em} FIG.}
\subfigure{\includegraphics[width=0.48\linewidth, height= 1.2 in]{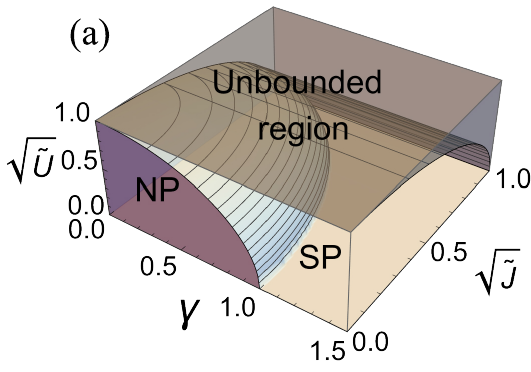}\label{fig:2(a)}}\hspace{-1 mm}
\subfigure{\includegraphics[width=0.48\linewidth, height= 1.2 in]{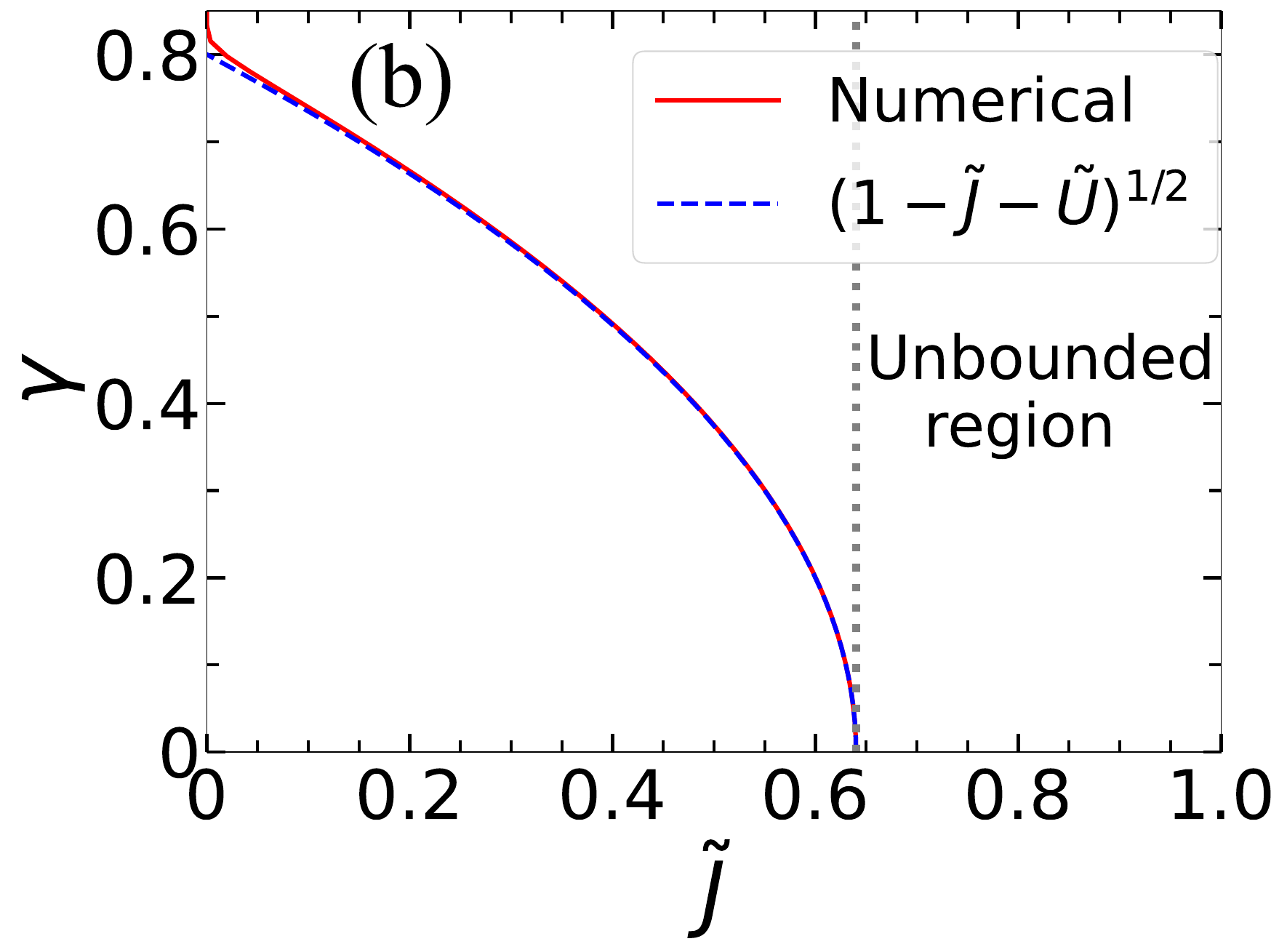}\label{fig:2(b)}}
\caption{\justifying \label{fig:2}(a) Phase diagram of the on-site effective Hamiltonian in the classical oscillator limit and at zero temperature, where the surface between the NP region and the SP region represents a $2nd$ order SPT. (b) Phase boundary of the on-site effective Hamiltonian of the model at zero temperature with $\tilde{U}=0.36$ and $\Omega/\omega=400$. The region below (above) the border represents the NP (SP). There is an unbounded region for $\tilde{J}>0.64$. The red line is obtained with the self-consistent method in Ref.~\citep{Schiro2013}. The dashed blue line is obtained with our method. They coincide with each other in the classical oscillator limit.}
\end{figure}
where 
\begin{equation}
H^{\mathrm{RS}}_{j}=\omega a^{\dag}_{j}a_{j}+(\frac{\Omega}{2}+U a^{\dag}_{j}a_{j})\sigma^{z}_j+g(a_{j}+a^{\dag}_{j})\sigma^{x}_{j}.
\label{eq:twenty}
\end{equation}

Here, $U$ and $J$ represent the strength of the Stark shift \citep{Grimsmo2013} and inter-cavity hopping, respectively. The stability condition of the system is $1-4J/\omega-U/\omega>0$, which will be discussed later. Employing the decoupling approximation \citep{Greentree2006} $a_{j}a^{\dag}_{j+1}\approx a_{j}\psi^{*} + a^{\dag}_{j+1}\psi-|\psi|^2$ where the coherent field $\langle a_{j}\rangle=\psi$ is the expectation value of $a_j$ in terms of the stable state, we obtain a decoupled Hamiltonian:
\begin{equation}
H\approx\sum^{M}_{j}{H^{\mathrm{eff}}_j},
\end{equation}
where
\begin{equation}
H^{\mathrm{eff}}_{j}=H^{\mathrm{RS}}_{j}-J[2(a_j+a^{\dag}_j)(\psi+\psi^{*})-2|\psi|^2-\psi^2-\psi^{*2}].
\label{eq:twenty-one}
\end{equation}
We focus on the on-site effective Hamiltonian $H^{\mathrm{eff}}_j$ and consider the classical oscillator limit $\eta=\Omega/\omega\rightarrow\infty$. The rescaled Hamiltonian reads
\begin{eqnarray}
H^{\mathrm{eff}}_j/\Omega &=&(\frac{1}{2}+\frac{U}{\omega} \frac{a^{\dag}_ja_j}{\eta})\sigma^{z}_j+\frac{g}{\sqrt{\Omega\omega}}(\frac{a_j}{\sqrt{\eta}}+\frac{a^{\dag}_j}{\sqrt{\eta}})\sigma^{x}_{j}\nonumber\\
& &+
\frac{a^{\dag}_ja_j}{\eta}-\frac{2J}{\omega}(\frac{a_j}{\sqrt{\eta}}+\frac{a^{\dag}_j}{\sqrt{\eta}})(\frac{\psi}{\sqrt{\eta}}+\frac{\psi^{*}}{\sqrt{\eta}})\nonumber\\
& &+
\frac{2J}{\omega}\frac{|\psi|^2}{\eta}+\frac{J}{\omega}(\frac{\psi^2}{\eta}+\frac{\psi^{*2}}{\eta}).
\end{eqnarray}

Applying the same method in the previous section, we obtain the Landau potential for $H^{\mathrm{eff}}_j$ 
\begin{eqnarray}
\phi(u, v)&=&-\frac{1}{2}\sqrt{[1+2\tilde{U}(u^2+v^2)]^2+4\gamma^2 u^2}\nonumber\\
& &+(1-\tilde{J})u^2+v^2,
\label{eq:twenty-two}
\end{eqnarray}
where we omit the subscript $j$ for convenience. Here, the coupling parameter settings are $\tilde{U}=U/\omega$, $\tilde{J}=4J/\omega$, and $\gamma=2g/\sqrt{\Omega\omega}$. The real and imaginary parts of $\psi/\sqrt{\eta}$ are denoted by $u$ and $v$, respectively. 

We next briefly discuss the stability of this model by studying the on-site Landau potential. Considering that the quartic terms of $u$ and $v$ with coefficient $\tilde{U}^2$ are leading-order within the square root in Eq.~\eqref{eq:twenty-two}, the stability of the system consequently requires $1-\tilde{J}-\tilde{U}>0$ for the Landau potential to have a lower bound. Due to the fact that the Landau potential Eq.~\eqref{eq:twenty-two} with specific $\vec{z}=(u, v)$ and $\vec{\lambda}=(\gamma, \tilde{J}^{1/2}, \tilde{U}^{1/2})$ satisfies Eq.~\eqref{landaup} formally, the phase diagram of this model is predicted to exhibit the above-mentioned general feature. Since the phase diagram is determined by the GMLP, we need to find the minima of the Landau potential, which satisfy 
\begin{align}
\frac{\partial\phi}{\partial u}&=2u\left\{1-\tilde{J}-\frac{[1+2\tilde{U}(u^2+v^2)]\tilde{U}+\gamma^2}{\sqrt{[1+2\tilde{U}(u^2+v^2)]^2+4\gamma^2 u^2}}\right\}=0,\label{eq:twenty-three}\\
\frac{\partial\phi}{\partial v}&=2v\left\{1-\frac{[1+2\tilde{U}(u^2+v^2)]\tilde{U}}{\sqrt{[1+2\tilde{U}(u^2+v^2)]^2+4\gamma^2 u^2}}\right\}=0. \label{eq:twenty-four}
\end{align}
Obviously, the GMLP is located at the origin $(u,v)=0$ when $|\vec{\lambda}|$ approaches zero. Considering $\tilde{U}<1-\tilde{J}$, the expression in the curly brace of Eq.~\eqref{eq:twenty-four} is always positive, meaning that there is no nonzero solution to Eq.~\eqref{eq:twenty-four}, so that we can neglect the imaginary part. The Landau potential can be rewritten as 
\begin{equation}
\phi(u)=(1-\tilde{J})u^2-\frac{1}{2}\sqrt{(1+2\tilde{U}u^2)^2+4\gamma^2 u^2}.
\label{eq:twenty-six}
\end{equation}  
Considering that it is still complicate to determine the minimum by taking the derivative, we choose to solve $\phi(u\neq 0)=\phi(0)$. We find that nonzero solutions exist for $\gamma^2+\tilde{J}+\tilde{U}>1$ under the stability condition. Hence the phase boundary is identified as $\gamma^2+\tilde{J}+\tilde{U}-1=0$. In addition, the expression of nonzero solutions reads
\begin{equation}
u^2=\frac{\gamma^2+\tilde{J}+\tilde{U}-1}{(1-\tilde{J}+\tilde{U})(1-\tilde{J}-\tilde{U})},
\label{eq:twenty-eight}
\end{equation} 
which tends to zero at the phase boundary, indicating a $2nd$ order SPT because the change of the order parameter $u^2$ is continuous from the NP to the SP. It is worth noting that the general phase diagram feature is well reflected in the boundary $|\vec{\lambda}|^2=1$. We draw the phase diagram by finding the GLMP of the Landau potential Eq.~\eqref{eq:twenty-six}. The general feature is straightforward, as shown in Fig.~\ref{fig:2(a)}. 

Another way to calculate the phase boundary without taking the classical oscillator limit is to determine $\psi=\langle a\rangle$ self-consistently \citep{Schiro2013}, where $\langle a \rangle $ is the mean value of $a$ in the ground state of the single site effective Hamiltonian Eq.~\eqref{eq:twenty-one}, which gives
\begin{equation}
\frac{1}{\tilde{J_{\mathrm{c}}}}=\omega\sum_{n}|\langle n|a+a^{\dag}|\mathrm{GS}\rangle|^2\frac{1}{E_{n}-E_{\mathrm{g}}},
\label{eq:twenty-nine}
\end{equation} 
where $|\mathrm{GS}\rangle$ and  $|n\rangle$ are the ground and excited states of the on-site Rabi-Stark Hamiltonian with energies $E_{\mathrm{g}}$ and $E_n$, respectively. As a function of $\gamma$, the right hand side of Eq.~\eqref{eq:twenty-nine} can be numerically obtained by exact diagonalization of the Rabi-Stark Hamiltonian. We compare this result with the classical oscillator mean field result and find that they are well consistent when $\eta\rightarrow\infty$, as shown in Fig.~\ref{fig:2(b)}. Therefore, our method works well in the classical oscillator limit.

\section{\label{sec:level5}Anisotropic Rabi-Stark model} 

Finally, we consider the anisotropic couplings. To investigate the effects of anisotropy on the phase diagram feature, we consider the anisotropic Rabi-Stark model in the classical oscillator limit for $U<\omega$. It should be pointed out that the quantum phase transitions of the anisotropic Rabi-Stark model have been studied by Xie \emph{et al.} \citep{Xie2020} at $|U|=\omega$ for the finite frequency ratio case. The Hamiltonian corresponds to the first four terms with $M=N=1$ in Eq.~\eqref{eq:one}, which reads 
\begin{eqnarray}
H&=&g_1(a^{\dag}\sigma^{-}+a\sigma^{+})+g_2(a\sigma^{-}+a^{\dag}\sigma^{+}) \nonumber \\
& &+ \omega a^{\dag}a+\left(\frac{\Omega}{2}+U a^{\dag}a\right)\sigma^{z},
\label{eq:thirty}
\end{eqnarray} 
where $g_1$ and $g_2$ are the coupling parameters for the rotating-wave term and counter-rotating term, respectively. The Hamiltonian in Eq.~\eqref{eq:thirty} possesses a $Z_2$ symmetry with parity operator $\Pi=\exp[\mathrm{i}\pi(a^{\dag}a+\sigma^{+}\sigma^{-})]$. Following the previous approaches developed, the Landau potential at zero temperature is derived as 
\begin{eqnarray}
\phi(u, v)&=&-\frac{1}{2}\sqrt{[1+2\tilde{U}(u^2+v^2)]^2+4\gamma^{2}_1 u^2+4\gamma^{2}_2 v^2}\nonumber\\
& &+u^2+v^2
,
\label{eq:thirty-one}
\end{eqnarray}
where $\gamma_1=(g_1+g_2)/\sqrt{\Omega\omega}$, $\gamma_2=(g_1-g_2)/\sqrt{\Omega\omega}$, and $\tilde{U}=U/\omega$ are the dimensionless coupling constants. Similarly, Eq.~\eqref{eq:thirty-one} has the form of Eq.~\eqref{landaup} if $\vec{z}=(u, v)$ and $\vec{\lambda}=(\gamma_1, \gamma_2, \tilde{U}^{1/2})$ are identified. Therefore, 
\begin{figure}[!htb]
\renewcommand\figurename{\hspace{1 em} FIG.}
\includegraphics[width=0.65\linewidth]{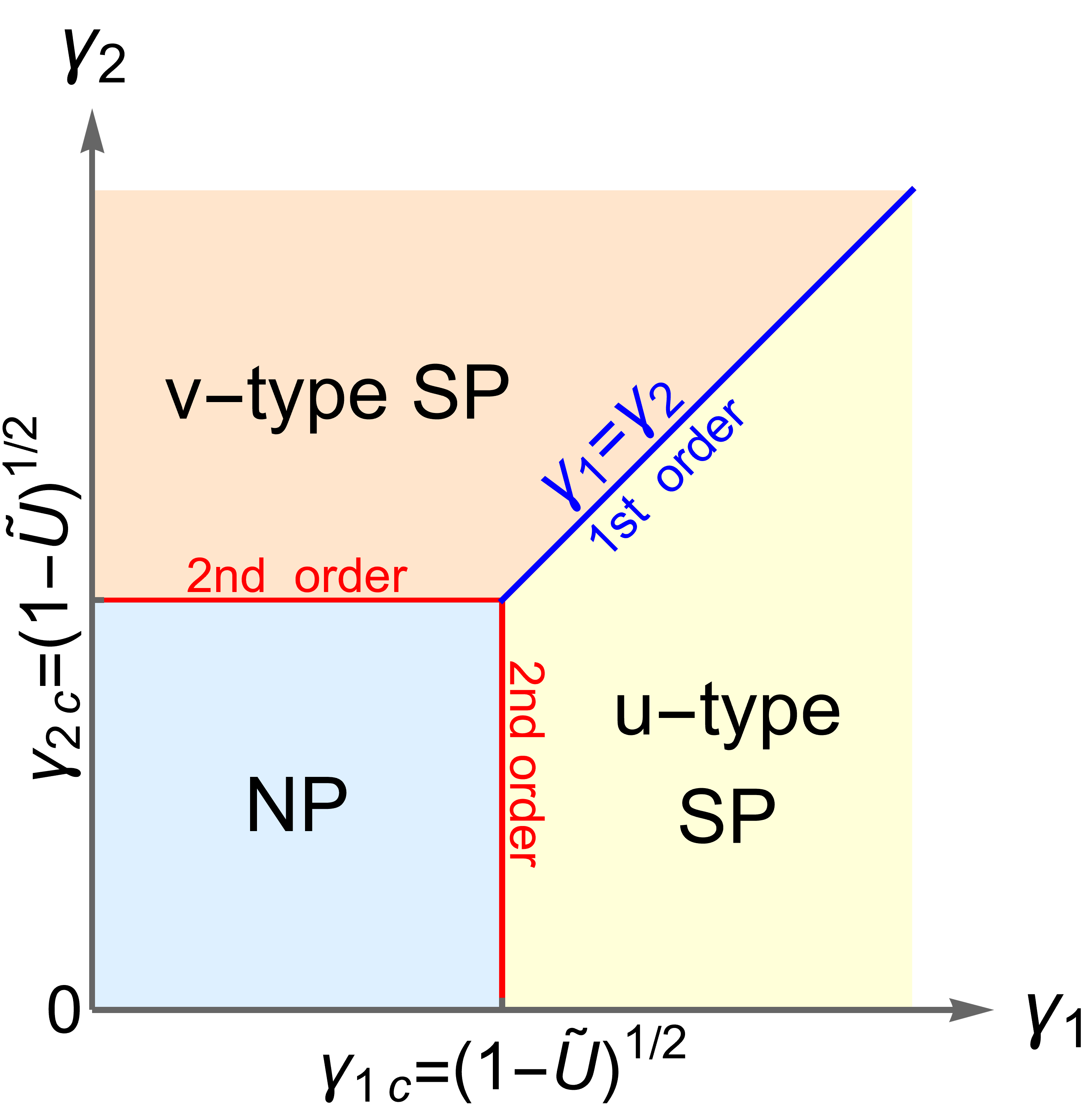}
\caption{\justifying \label{fig:3}Phase diagram of the model in the classical oscillator limit $\Omega/\omega\rightarrow\infty$ at zero temperature.}
\end{figure} the general phase diagram feature still holds. In addition, the stability condition $\tilde{U}<1$ should be satisfied to ensure that the Landau potential is bounded from below. Subsequently, we start to discuss the minima of the Landau potential
\begin{align}
\frac{\partial\phi}{\partial u}&=2u\left\{1\!-\!\frac{[1+2\tilde{U}(u^2+v^2)]\tilde{U}+\gamma^2_1}{\sqrt{[1\!+\!2\tilde{U}(u^2\!+\!v^2)]^2\!+\!4(\gamma^2_1 u^2\!+\!\gamma^2_2 v^2)}}\right\}\!=\!0,\label{eq:thirty-two}\\
\frac{\partial\phi}{\partial v}&=2v\left\{1\!-\!\frac{[1+2\tilde{U}(u^2+v^2)]\tilde{U}+\gamma^2_2}{\sqrt{[1\!+\!2\tilde{U}(u^2\!+\!v^2)]^2\!+\!4(\gamma^2_1 u^2\!+\!\gamma^2_2 v^2)}}\right\}\!=\!0. \label{eq:thirty-three} 
\end{align}
Apparently, $\phi(0,0)$ is the global minimum as $|\vec{\lambda}|$ is close to zero. The ratio of $\gamma_1$ to $\gamma_2$ determines which equation has nonzero solutions. When $\gamma_1=\gamma_2=\gamma$, Eqs.~\eqref{eq:thirty-two} and \eqref{eq:thirty-three} become completely identical in form. It follows that $u$ and $v$ equally contribute to the rescaled mean photon number. 

\begin{figure}[!htb]
\centering
\renewcommand\figurename{\hspace{1 em} FIG.}
\subfigure{\includegraphics[width=0.48\linewidth, height= 1.15 in]{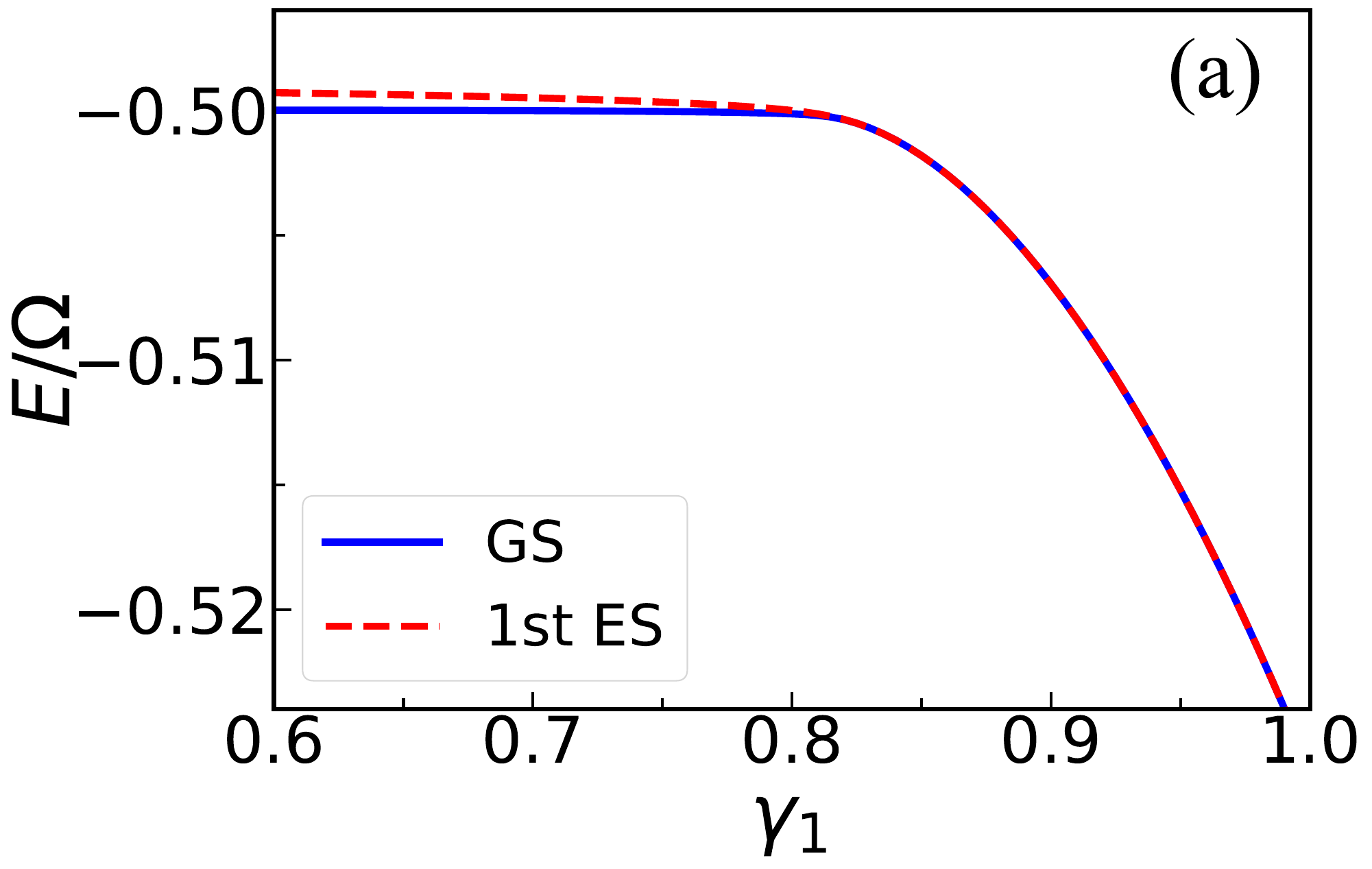}\label{fig:4(a)}}\hspace{-1 mm}
\subfigure{\includegraphics[width=0.48\linewidth, height= 1.2 in]{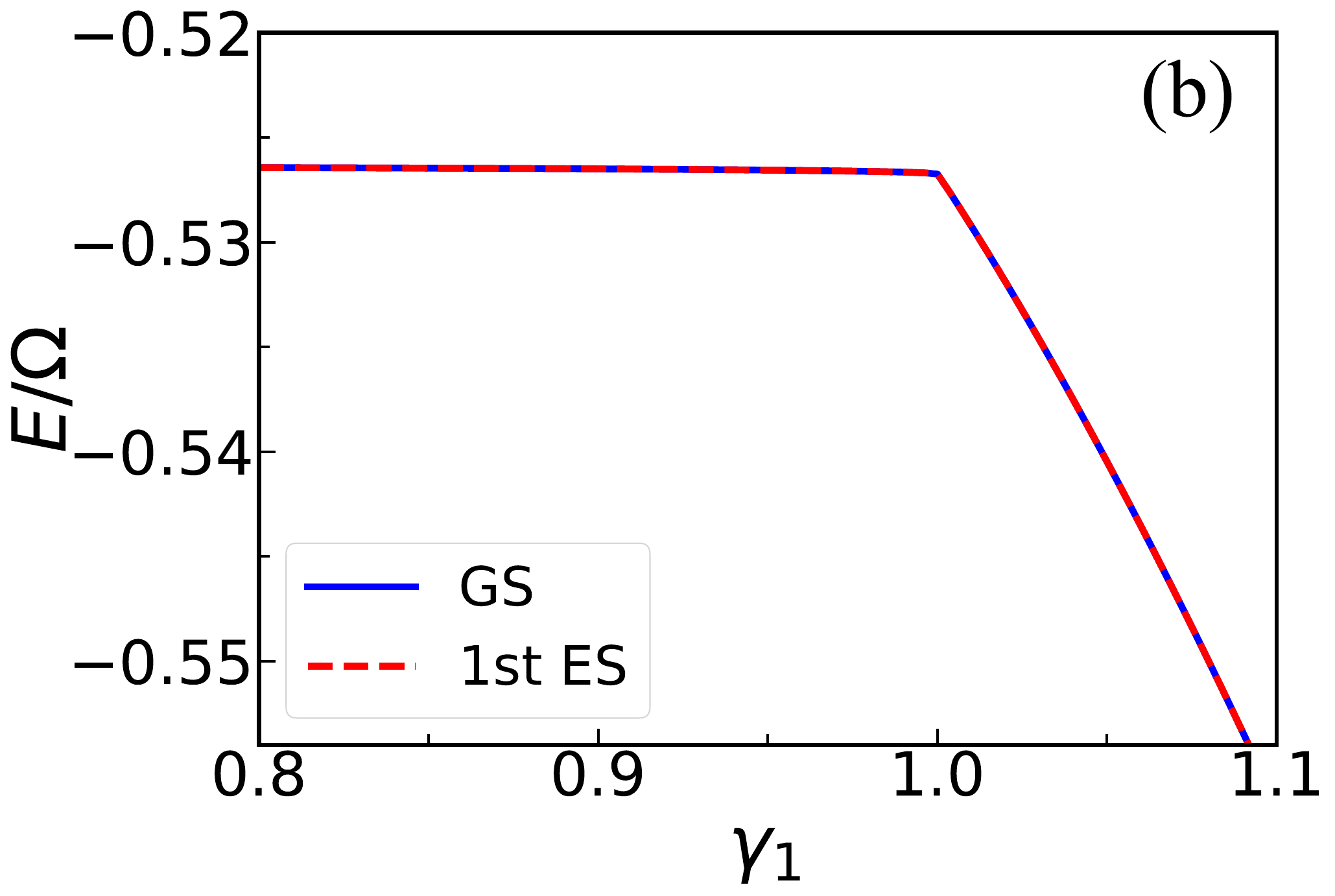}\label{fig:4(b)}}\vskip -8 pt
\subfigure{\includegraphics[width=0.48\linewidth, height= 1.15 in]{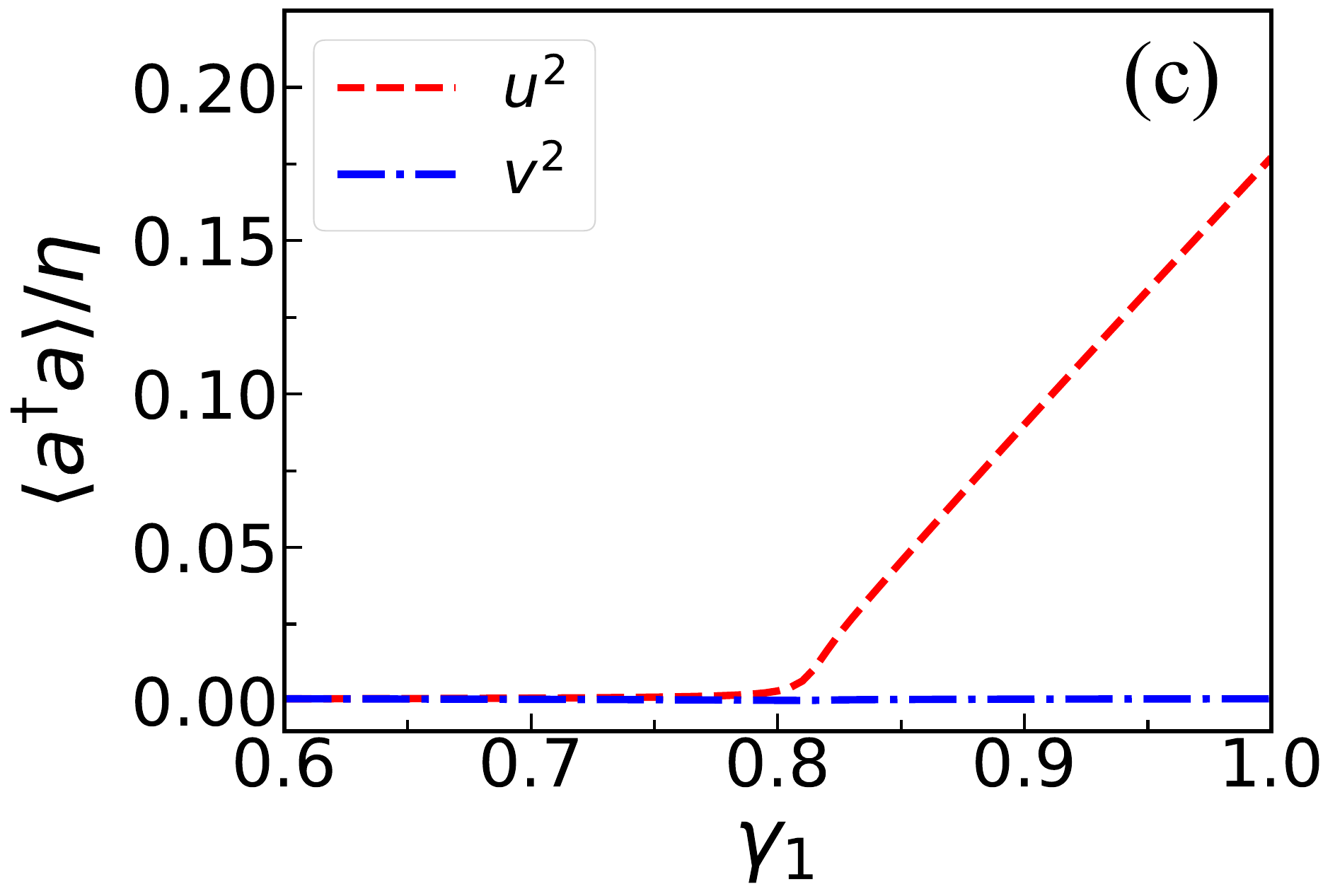}\label{fig:4(c)}}\hspace{-1 mm}
\subfigure{\includegraphics[width=0.48\linewidth, height= 1.2 in]{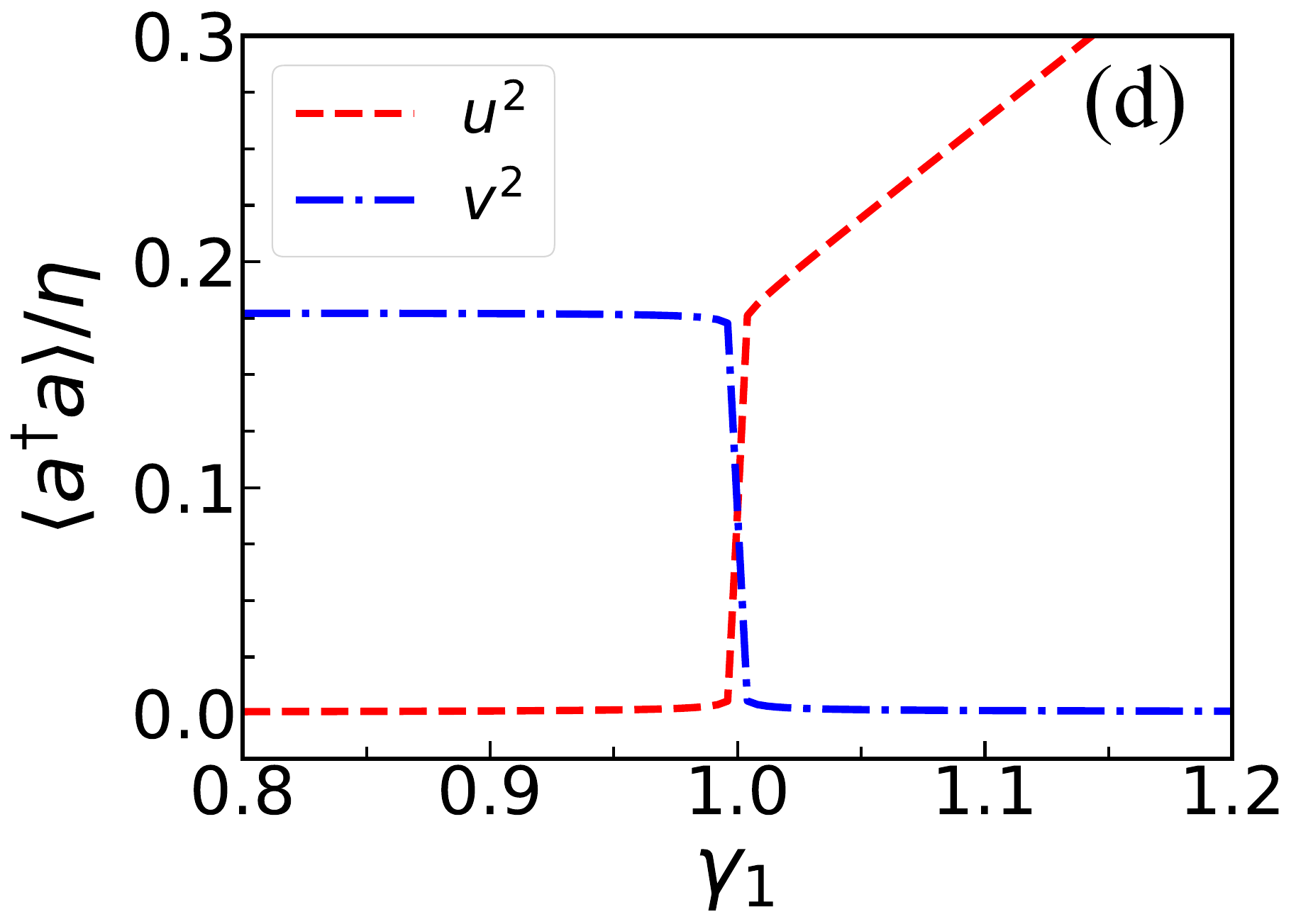}\label{fig:4(d)}}
\caption{\justifying \label{fig:4}The energies in the ground state (GS) and the first excited state (1st ES) as a function of $\gamma_1$ for (a) $\gamma_2=0.6$, (b) $\gamma_2=1$. The rescaled mean photon number in the ground state as a function of $\gamma_1$ for (c) $\gamma_2$=0.6, (d) $\gamma_2$=1. Here, $\langle a^{\dag}a\rangle/\eta=u^2+v^2$, $\Omega/\omega=400$ and $\tilde{U}=0.36$.}
\end{figure}

By solving $\phi(u, v)=\phi(0, 0)$, we find that nonzero solutions
\begin{equation}
u^2+v^2=\frac{\gamma^2+\tilde{U}-1}{(1-\tilde{U})(1+\tilde{U})}
\label{thirty-two}
\end{equation} 
exists when $\gamma^2+\tilde{U}-1>0$. This indicates a $2nd$ order SPT because the change of the order parameter $u^2+v^2$ is continuous at the phase boundary $\gamma^2+\tilde{U}-1=0$, i.e., $|\vec{\lambda}|=1$. The boundary can be reduced to that in the Jaynes-Cummings model if $\tilde{U}=0$. In the same way, we obtain the phase boundaries $\gamma^2_1+\tilde{U}-1=0$ for $\gamma_1/\gamma_2>1$ and $\gamma^2_2+\tilde{U}-1=0$ for $\gamma_2/\gamma_1>1$. And we find two distinct $Z_2$ symmetry breaking SPs, each of which has a rescaled mean photon number in the ground state depending solely on $u$ or $v$. We call them $u$-type SP and $v$-type SP, respectively. We also find that the phase transitions from the NP to both SPs are $2nd$ order while the quantum phase transition between two SPs is $1st$ order. Based on the analysis above, we draw the phase diagram of the model in the ($\gamma_1$, $\gamma_2$) plane in Fig.~\ref{fig:4}. The phase diagram has a similar structure to that in Ref.~\citep{Baksic2014}, with the difference that the critical coupling values are associated with $\tilde{U}$ here. It is worth pointing out that despite the presence of two SPs, the general phase diagram feature of the system entering the SP from the NP radially in the coupling parameter space is still preserved. 
To confirm the validity of our analysis on the phase diagram, we numerically calculate the low energy spectrum and the rescaled mean photon number in the ground state at certain values of $\tilde{U}$ and $\gamma_2$, as shown in Fig.~\ref{fig:4}. In Fig.~\ref{fig:4(a)}, the ground state and the first excited state are not approximately degenerate until $\gamma_1$ reaches the $2nd$ order phase transition point $\gamma_{1c}=0.8$. The appearance of the degeneracy implies the $Z_2$ symmetry breaking. The continuous change of the $u$-dependant rescaled mean photon number at the critical point indicates that the transition from the NP to the $u$-type SP is a $2nd$ order SPT, as shown in Fig.~\ref{fig:4(c)}. In Fig.~\ref{fig:4(b)}, since $\gamma_2=1>\gamma_{2c}$, the system is always in the SP and this can be verified by the degeneracy between two lowest energy states. Moreover, the sharp decrease of the ground state energy in the limit of $\eta\rightarrow\infty$ lead to the discontinuity of its first-order derivative, which signals the emergence of a $1st$ order quantum phase transition.  The abrupt changes of $u^2$ and $v^2$ indicate that system undergoes a $1st$ quantum phase transition from the $v$-type SP to the $u$-type SP in Fig.~\ref{fig:4(d)}. 

To this end, we confirm the general light-matter interaction model Eq.~\eqref{eq:one} has the phase diagram feature we found, since the last line regarding qubit-qubit interaction generally will not change the general form of Landau potential Eq.~\eqref{landaup}.
 
\section{\label{sec:level6}Conclusions}

We find a general phase diagram feature of light-matter systems that by specifically choosing a coupling parameter vector $\vec{\lambda}$, the systems are in NP at the origin of the coupling parameter space, and they go to SP radially and stay there afterwards. We can obtain the phase boundary and determine the rescaled mean photon number there with a concise method, so that one can identify whether SPT exists and is first-order or second-order. We illustrate this result with the multimode Dicke model with both one- and two-photon terms, the Rabi-Stark-Hubbard model, and the anisotropic Rabi-Stark model. We find the addition of photon modes will reduce the coupling strength requirement on each mode for SPT, even to the strong coupling regime. We also find disorder will shift the phase boundary. Our work provides insights into SPT phase diagrams, enabling advanced applications in quantum information processing.

\section*{Acknowledgements}

This work was supported by the Scientific Research Fund of Hunan Provincial Education Department (Grant No. 23A0135), Natural Science Foundation of Hunan Province, China (Grants No. 2024JJ10045 and No. 2022JJ30556),
Guizhou Provincial Basic Research Program (Natural Science) (Grant No. ZK[2024]021). National Natural Science Foundation of China (Grant No.~11704320). 
\appendix*
\section{Landau potential for general light-matter systems}

\section{The stability condition of multimode Dicke model with both one- and two-photon terms}

The stability condition can be determined by ensuring that the Landau potential is lower bounded. Here, we apply a brief analysis to the Landau potential in Eq.~\eqref{eq:nine} in the main text. The Landau potential at zero temperature is rewritten as
\begin{equation}
\phi(\vec{z})=\sum^{M}_{\nu=1}{(u^2_{\nu}+v^2_{\nu})}-\frac{1}{2}\sqrt{1+\xi^2},
\end{equation}
with $\xi=2\sum^{M}_{\nu=1}{[\gamma_{\nu}u_{\nu}+\gamma^{\prime}_{\nu}(u^2-v^2)]}$. To consider the asymptotic behavior at $u,v\gg 1$, we approximate the Landau potential as
\begin{equation}
\phi(\vec{z})_{\pm}=\sum^{M}_{\nu=1}{\phi(u_{\nu}, v_{\nu})_{\pm}},
\end{equation}
with 
\begin{eqnarray}
\phi(u_{\nu}, v_{\nu})_{\pm}&=&(1\mp\gamma^{\prime}_{\nu})\left[u\mp\frac{\gamma_{\nu}}{2(1\mp\gamma^{\prime}_{\nu})}\right]^2-\frac{\gamma^2_{\nu}}{4(1\mp\gamma^{\prime}_{\nu})}\nonumber\\
& &+
(1\pm\gamma^{\prime}_{\nu})v^2_{\nu}.
\end{eqnarray}
The subscripts of $\phi(u_{\nu}, v_{\nu})_{\pm}$ depend on whether $\xi$ takes the positive or negative sign. The Landau potential $\phi(u_{\nu}, v_{\nu})_{\pm}$ is quadratic and the condition $\gamma^{\prime}_{\nu}<1$ must be required to ensure that $\phi(u_{\nu}, v_{\nu})_{\pm}$ is bounded from below both in the $u$ and $v$ plane. 
%

\end{document}